\documentclass[traditabstract]{aa}
\usepackage{graphicx}
\usepackage{txfonts}
\usepackage{color}

\newcommand{\obj}{H1413+117}

\begin{document}
\title{Polarization microlensing in the quadruply imaged broad
absorption line quasar H1413+117\thanks{Based on observations made
with ESO Telescopes at the Paranal Observatory (Chile). ESO program
ID: 386.B-0337}}
\author{D. Hutsem\'ekers\inst{1,}\thanks{Senior Research Associate F.R.S.-FNRS}, 
        D. Sluse\inst{1},
        L. Braibant\inst{1,}\thanks{Research Assistant  F.R.S.-FNRS},
        T. Anguita\inst{2,3}
        }
\institute{
    Institut d'Astrophysique et de G\'eophysique,
    Universit\'e de Li\`ege, All\'ee du 6 Ao\^ut 19c, B5c, 4000
    Li\`ege, Belgium
    \and 
    Departamento de Ciencias Fisicas, Universidad Andres Bello,
    Fernandez Concha 700, Las Condes, Santiago, Chile
    \and 
    Millennium Institute of Astrophysics, Chile
    }
\date{Received ; accepted: }
\titlerunning{Polarization microlensing in H1413+117} 
\authorrunning{D. Hutsem\'ekers et al.}
\abstract{We have obtained spectropolarimetric observations of the
four images of the gravitationally lensed broad absorption line quasar
H1413+117.  The polarization of the microlensed image D is
significantly different, both in the continuum and in the broad lines,
from the polarization of image A, which is essentially unaffected by
microlensing.  The observations suggest that the continuum is
scattered off two regions, spatially separated, and producing roughly
perpendicular polarizations. These results are compatible with a model
in which the microlensed polarized continuum comes from a compact
region located in the equatorial plane close to the accretion disk and
the non-microlensed continuum from an extended region located along
the polar axis.
}
\keywords{Gravitational lensing -- Quasars: general -- Quasars:
absorption lines --Quasars: individual : H1413+117}
\maketitle
%
%
%
\section{Introduction}
\label{sec:intro}

Spectropolarimetry is a powerful tool to study and disentangle the
inner regions of quasars, otherwise impossible to resolve
spatially. Most often optical linear polarization originates
from scattering off a separate region that acts as a mirror. In
this kind of situation, the separation of the polarized light from the
unpolarized light usually offers an additional line of sight to
different quasar regions, providing information about the geometry and
location of quasar components, such as the source of continuum or the
broad emission/absorption line regions (e.g., Antonucci and Miller
\cite{ant85}; Smith et al. \cite{smi04}; Kishimito et
al. \cite{kis08}).

Gravitational microlensing, on the other hand, can selectively magnify
quasar inner regions, such as the accretion disk or even part of the
broad line region, in one of the lensed images.  A comparison with
unaffected images can provide constraints on the size, location, and
kinematics of the quasar inner regions (e.g., Eigenbrod et
al. \cite{eig08}; Blackburne et al. \cite{bla11}; Sluse et
al. \cite{slu11}; Braibant et al. \cite{bra14}).

Broad absorption lines (BALs) are seen in approximately 20\%
of optically selected quasars (Knigge et al. \cite{kni08}). BAL
quasars are characterized by deep blueshifted absorption in the
resonance lines of ionized species revealing fast and massive outflows,
which likely affect the evolution of the host galaxy and
intergalactic medium (Silk and Rees \cite{sil98}).  Spectropolarimetry
has proved useful to constrain the geometry and dynamics of the
outflows, which are usually attributed to roughly equatorial winds
launched from the accretion disk (Goodrich and Miller \cite{goo95};
Ogle et al. \cite{ogl99}; Lamy and Hutsem\'ekers \cite{lam04}; Young
et al. \cite{you07}), although polar winds are observed in some
radio-loud BAL quasars (Zhou et al. \cite{zho06}; Ghosh and Punsly
\cite{gho07}; DiPompeo et al. \cite{dip13}; Bruni et
al. \cite{bru13}).  In one particular case, microlensing has helped
to disentangle the intrinsic absorption line from the emission line
profile, providing evidence for outflows with both equatorial and
polar components.  (Hutsem\'ekers et al. \cite{hut10}, hereafter
Paper~I).

Here, we report on the first spectropolarimetric observations of the four
images of a gravitationally lensed quasar, the quadruply imaged broad
absorption line quasar H1413+117, known to be both polarized and
microlensed (Angonin et al. \cite{ang90}; Hutsem\'ekers \cite{hut93};
{\O}stensen et al. \cite{ost97}; Chae et al. \cite{cha01}; Anguita et
al. \cite{ang08}; Paper~I; O'Dowd et al. \cite{odo15}).  In Sluse et
al. (\cite{slu15}, hereafter Paper~II), we have unveiled the existence
of two sources of continuum, especially a spatially separated
or extended continuum source, using differences induced by
microlensing in the total flux spectra.  The present paper focuses on
the analysis of the polarization of the individual images.  Broadband
polarization measurements of the four images were already reported in
Chae et al. (\cite{cha01}) and in Paper~I, with inconclusive results
(see also Hales and Lewis \cite{hal07}).

\section{Observations and data reduction}
\label{sec:data}

\begin{figure*}[t]
\resizebox{\hsize}{!}{\includegraphics*{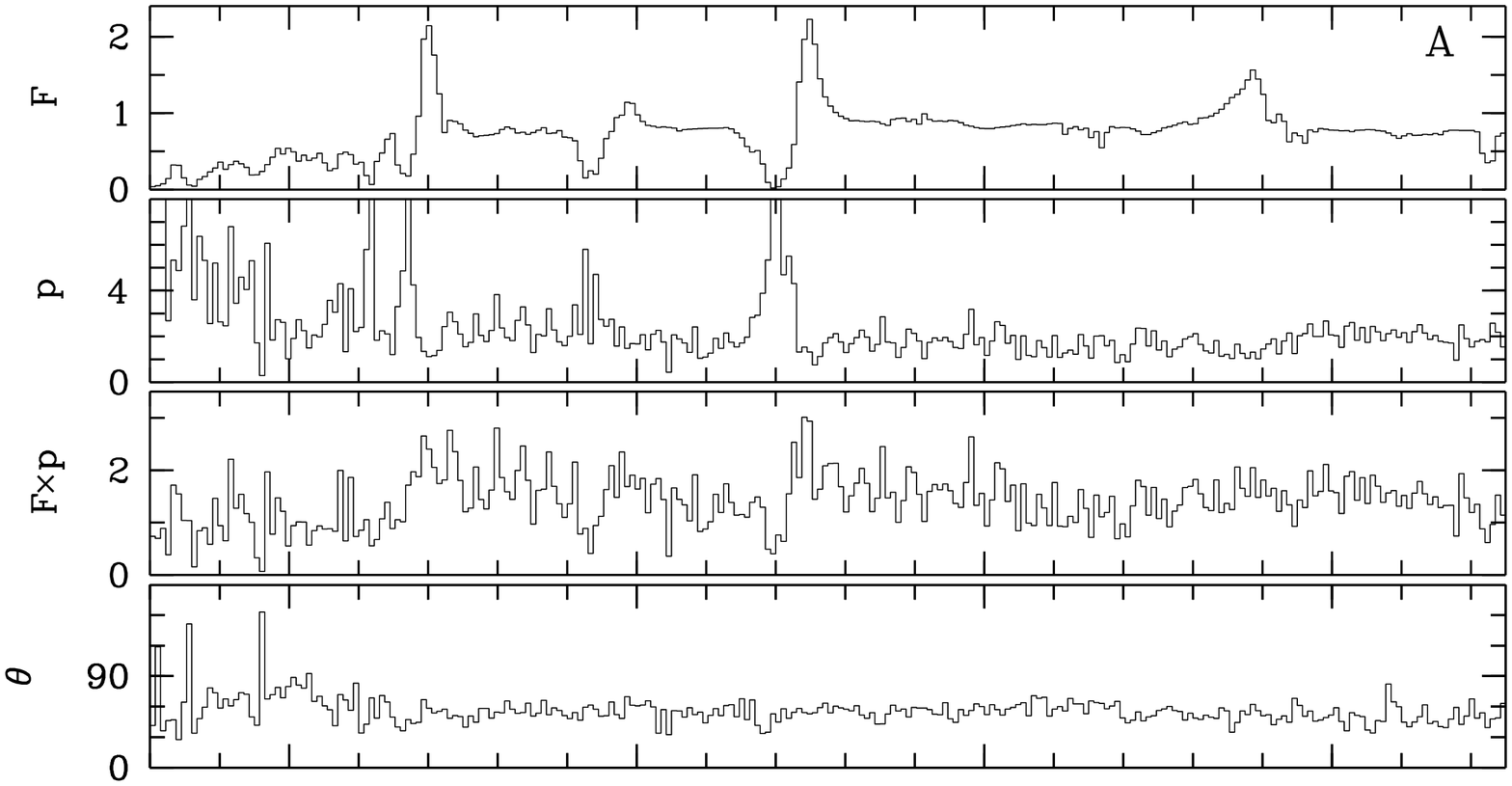}\includegraphics*{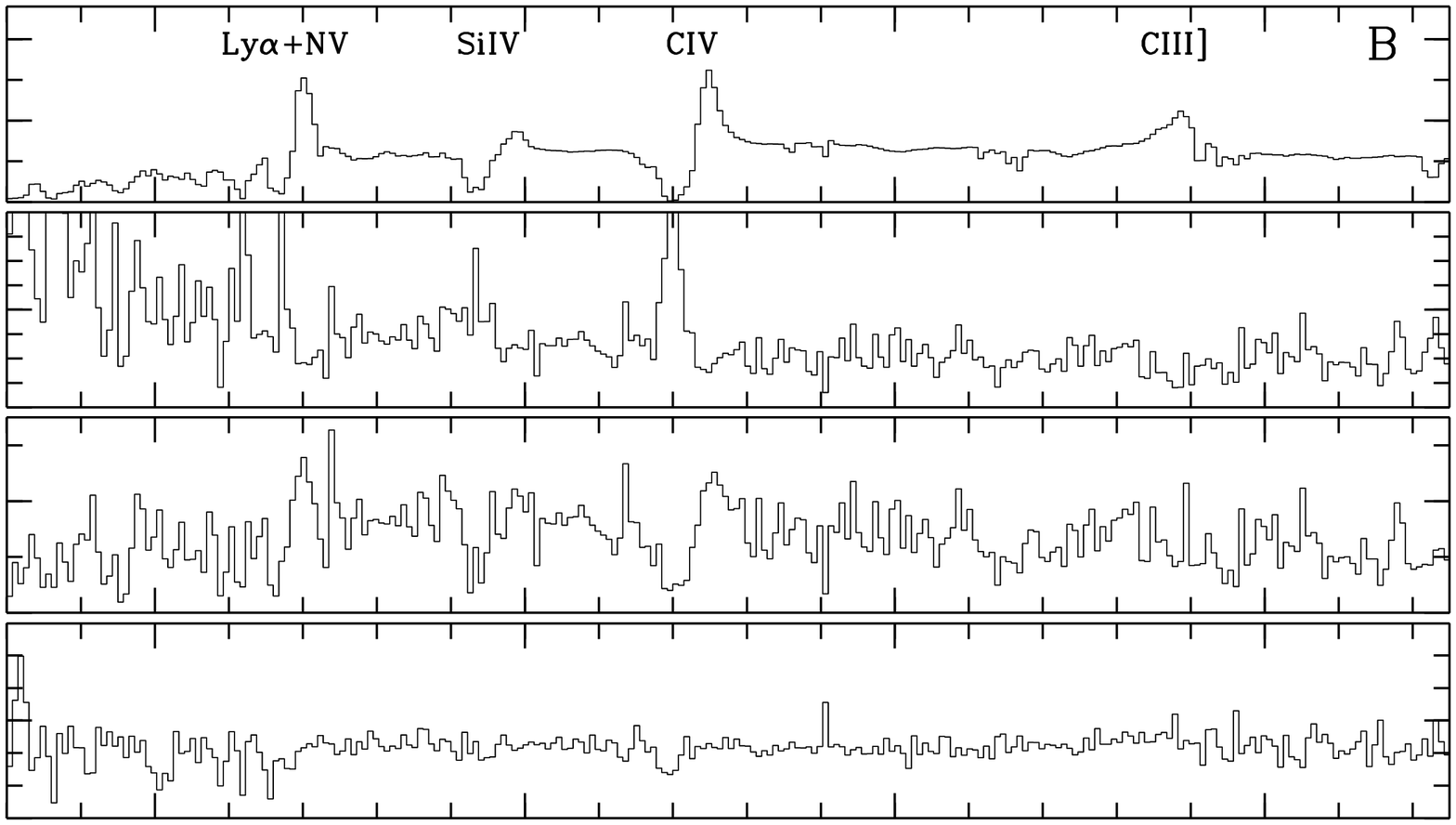}}
\resizebox{\hsize}{!}{\includegraphics*{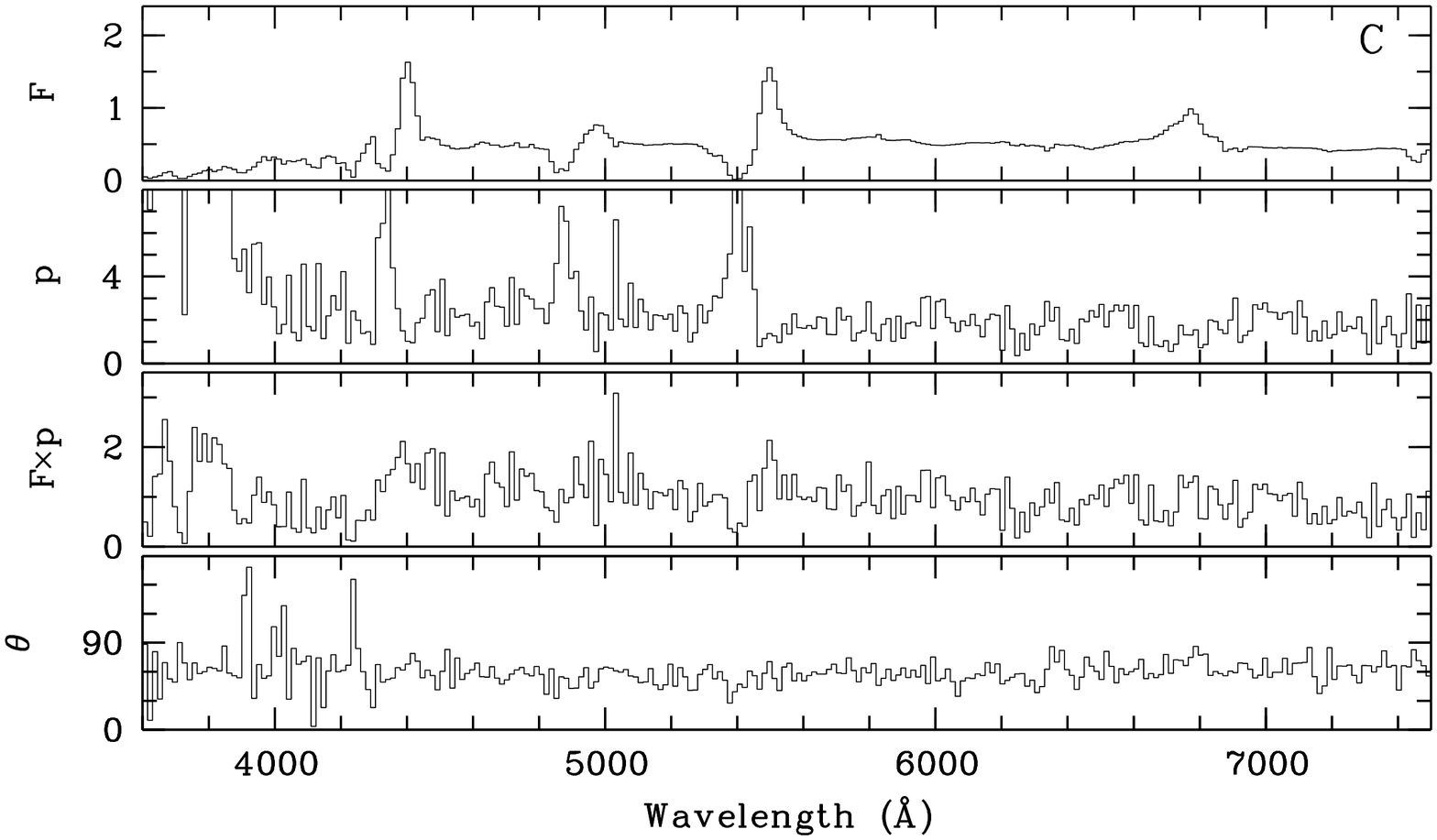}\includegraphics*{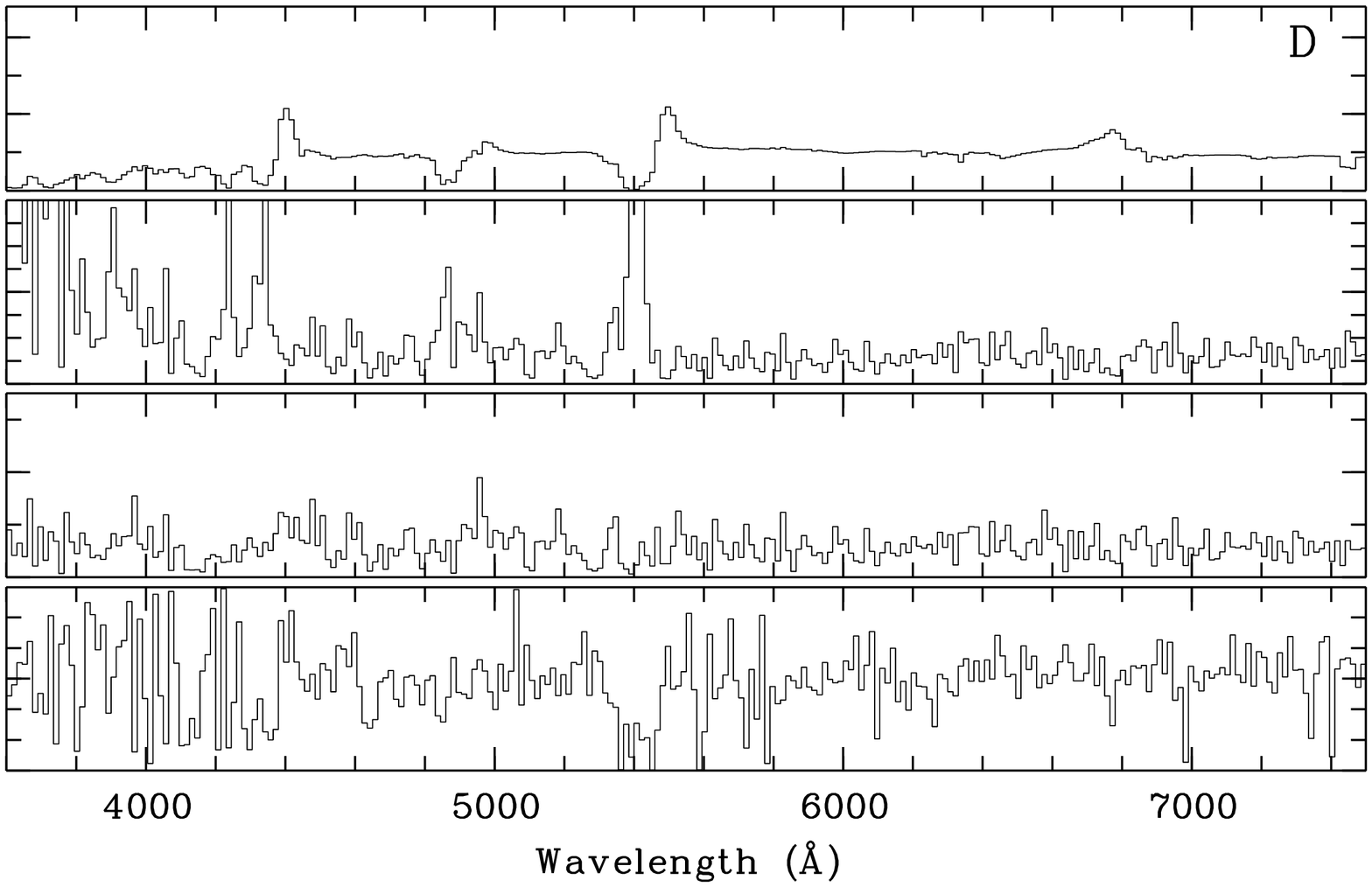}}
\caption{The total flux $F(\lambda)$, (in arbitrary units), the linear
polarization degree $p(\lambda)$, (in \%), the polarized flux
$F(\lambda)\times p(\lambda)$, (in arbitrary units), and the polarization
position angle $\theta(\lambda)$, (in degree, east of north) for images
A, B, C, D of H1413+117.}
\label{fig:spola}
\end{figure*}

The observations were carried out in March-April 2011 with the ESO
Very Large Telescope equipped with the FORS2 instrument in its linear
spectropolarimetry mode (FORS User Manual, VLT-MAN-ESO-13100-1543).
The spectra were obtained by positioning a 0\farcs4-wide MOS slit
through either the A-D or the B-C images of H1413+117 (naming
convention from Paper I). Grisms 300V and 300I were used to cover the
0.36-0.90 $\mu$m and 0.60-1.03 $\mu$m spectral ranges,
respectively. Each observation was carried out twice with a seeing better
than 0\farcs6 (cf. Paper~II, for other details).  Exposures of
660s each were secured with the half-wave plate rotated at the
position angles 0\degr, 22.5\degr, 45\degr , and 67.5\degr.  For each
image and grism, 16 spectra (two orthogonal polarizations, four half-wave
plate angles, two epochs) were obtained and extracted by fitting Moffat
profiles along the slit.

For each image of \obj , the normalized Stokes parameters $q(\lambda)$
and $u(\lambda)$, the linear polarization degree $p(\lambda)$, the
polarization position angle $\theta(\lambda)$, the total flux
$F(\lambda)$, and the polarized flux $F(\lambda) \times p(\lambda)$
were computed from the individual spectra according to standard
recipes (e.g., FORS manual). To increase the signal to noise
ratio, the original spectra were rebinned over 15\AA\ wavelength
($\sim$9 pixel) bins and the two epochs added before computing the
polarimetric data. The spectra were corrected for the retarder plate
zero angle given in the FORS manual. The final blue spectra are
illustrated in Fig.~\ref{fig:spola}. Contamination by interstellar
polarization in our Galaxy is negligible for \obj\
(Hutsem\'ekers et al. \cite{hut98}; Ogle et al. \cite{ogl99}).

To enable comparison with previous broadband polarization
measurements, we integrated the spectra over the FORS2 V~High filter
and measured the broadband polarizations reported in
Table~\ref{tab:pola}.  The quoted uncertainties are estimated from the
differences between the measurements obtained at the two epochs
(separated by at most 26 days), which are in excellent
agreement.  Since the \ion{C}{iv} line appears in the V filter, we
also measured the polarization by integrating over a continuum region
excluding \ion{C}{iv}, namely the 0.57-0.64~$\mu$m wavelength
range. The difference between these two sets of broadband
measurements is negligible. Finally, polarization integrated over the
FORS2 I Bessel filter is also given, using both the 300V and 300I data
sets.

\section{The polarization of the four images}

\subsection{Overview}

The broadband measurements confirm the variability in both degree and
angle of the polarization in \obj, which was first reported by Goodrich and
Miller (\cite{goo95}).

In 2011, there was a net decrease of the continuum polarization degree
of image D with respect to the other images and previous epochs
(Table~\ref{tab:pola}).  The polarization angle measured for image D
also appears significantly different, by $\sim$25\degr , from the
angles measured in the other images. This difference is roughly
constant with time, apparently superimposed on intrinsic variations
that affect all images.

In agreement with previous spectropolarimetric observations (Goodrich
and Miller \cite{goo95}; Ogle et al. \cite{ogl99}; Lamy and
Hutsem\'ekers \cite{lam04}), we find a strong increase of the
polarization degree in the BAL troughs.  The strongest,
low-velocity absorption component is present but shallower in
polarized flux than in total flux.  These polarization properties are
usually interpreted by electron or dust scattering of the continuum,
where the scattered light is less absorbed than the direct light.

For images A, B, and C, a small rotation of the polarization angle is
observed in the BAL troughs, in particular \ion{C}{iv}, suggesting the
presence of two polarization sources with different polarization
angles.  The \ion{N}{v}, the \ion{C}{iv}, and possibly the \ion{Si}{iv}
emission lines appear in the polarized flux, but they are less
polarized than the continuum.  Although the polarization data of image
D are noisier, the \ion{C}{iv} absorption line is seen in the
polarized flux with no significant emission line. In that image, a
strong rotation of the polarization angle is observed across a large
part of the \ion{C}{iv} line profile.

\subsection{The continuum polarization: Analysis of the (D,A)
pair}

We focus on the (D,A) pair of images because at rest-frame optical-UV
wavelengths, image D is strongly microlensed while only image
A is not (Paper~II).  The fact that the polarization angle of the
continuum of image D is rotated with respect to the polarization angle
of the continuum of image A requires two sources of polarized
continuum with different polarization angles, one which is
microlensed and the other not\footnote{Differential extinction is
not observed between A and D. Interstellar polarization in the lens
can thus not explain the observed difference between A and D and more
particularly the variable polarization in D (Chae et
al. \cite{cha01}).  On the other hand, dust extinction is observed in
image B (Paper~I) and could contribute to the small systematic
polarization difference observed between A and B.}.  Because its
magnification decreases with increasing wavelength, the microlensed
continuum source was identified in Paper~II as the accretion disk.
This microlensed polarized continuum then originates from a compact
source, presumably disk-like, close to the accretion disk or from the
accretion disk itself\footnote{Chae et al. (\cite{cha01}) assumed on
the contrary that a part of the scattering region is microlensed while
the inner continuum-producing region (i.e., the accretion disk) is
not. This kind of scenario would not produce the observed chromatic
magnification of the continuum. Moreover, magnifying the continuum in
D by a factor $\sim$ 2 would be difficult if the caustic is offset
from the central source so as to magnify only a part of the scattered
continuum.}.

\begin{table}[t]
\caption{Broadband polarization of the four images of \obj }
\label{tab:pola}
\begin{tabular}{lcccc}\hline\hline \\[-0.10in]
Epoch / Band    &  A &  B  &  C & D \\ 
\hline \\[-0.10in]
& &   $p  \;(\%)$  & & \\ \hline \\[-0.10in]
1999 / F555W & 1.6$\pm$0.5&2.3$\pm$0.5&1.8$\pm$0.5&2.9$\pm$0.5\\
2008 / V & 1.4$\pm$0.1&2.4$\pm$0.1&1.2$\pm$0.1&2.0$\pm$0.1\\
2011 / V & 1.7$\pm$0.1&2.2$\pm$0.1&1.9$\pm$0.1&0.7$\pm$0.1\\
2011 / Vc& 1.7$\pm$0.1&2.0$\pm$0.1&1.7$\pm$0.1&0.8$\pm$0.2\\
2011 / I & 1.7$\pm$0.3&1.7$\pm$0.3&1.5$\pm$0.2&1.0$\pm$0.2\\
\hline \\[-0.10in]
& &  $\theta \;(\degr)$  & &  \\ \hline \\[-0.10in]
1999 / F555W & 75$\pm$9&65$\pm$6&71$\pm$8&102$\pm$5\\
2008 / V & 72$\pm$2&79$\pm$1&69$\pm$3&96$\pm$2\\
2011 / V & 55$\pm$2&65$\pm$2&57$\pm$2&82$\pm$4\\
2011 / Vc& 56$\pm$2&63$\pm$2&57$\pm$2&86$\pm$8\\
2011 / I & 47$\pm$5&64$\pm$5&55$\pm$4&78$\pm$6\\
\hline\\[-0.2cm]
\end{tabular}\\
\footnotesize{The linear polarization degree $p$ is given in percent
and the polarization position angle $\theta$ in degree, east of
north. The data obtained in 2008 and 1999 are from Paper~I and from
Chae et al.~(\cite{cha01}), respectively.  For the 2011 data, V and I
refer to measurements integrated over the FORS2 V and I filters, and
Vc over the 0.57 - 0.64~$\mu$m wavelength range.}
\end{table}

Under the hypothesis that the observed continuum polarization is the
sum of the polarization from a microlensed compact source and a
non-microlensed extended one, we can extract the intrinsic
polarizations of the two sources, $p_{c}$, $\theta_{c}$, $p_{e}$,
$\theta_{e}$, as detailed in Appendix A.  The results are listed in
Table~\ref{tab:polamod1d}. They were derived from the broadband
measurements using the magnification factor $\mu$ =2.0 assumed
constant with time (Paper~II).  Three values of $F_{e} / F_{c}$, the
flux ratio between the extended and compact continua, are considered.

\begin{figure}[t]
\resizebox{\hsize}{!}{\includegraphics*{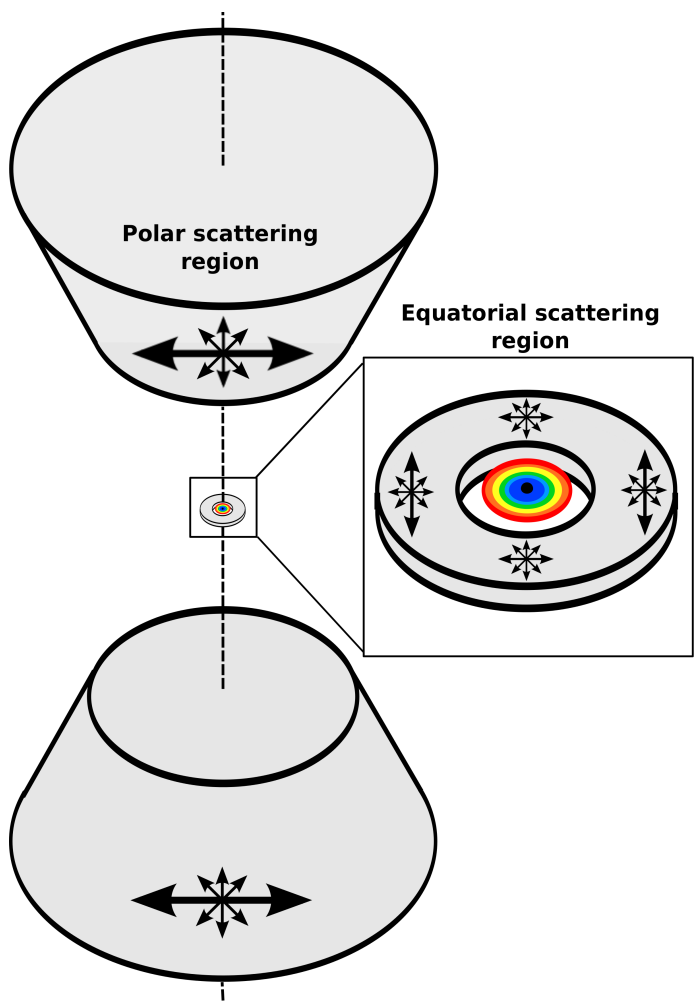}}
\caption{Schematic picture (not to scale; adapted from Smith et
al. 2004) illustrating the model adopted for interpreting the quasar
continuum polarization.  The equatorial scattering region is compact
and magnified by microlensing together with the central continuum
source (the accretion disk), while the polar scattering region is more
extended and not magnified.  The system axis is inclined with respect
to the line of sight to the observer. The arrows indicate the
direction of the resulting polarization projected onto the plane of
the sky for each scattering region. The equatorial scattering region
produces low polarization predominantly parallel to the system
symmetry axis, while the polar scattering region produces
high polarization perpendicular to the system axis.}
\label{fig:model}
\end{figure}

Although differences exist, likely due to the oversimplified model
considered here, the results appear remarkably consistent for the
different epochs and wavelength bands. The polarization of the compact
source is around $p_{c} \sim 3 \%$ and $\theta_{c} \sim 115 \degr$.
The polarization of the extended source is around $p_{e} \sim 10 \%$
and $\theta_{e} \sim 35 \degr$.  The polarizations appear
roughly orthogonal. They can be produced by electron or dust
scattering (Brown and McLean \cite{bro77}; Goosmann and Gaskell
\cite{goo07}).  These results are consistent with a single central
source of continuum, the accretion disk, which is scattered off a
compact low-polarization region close to it in the equatorial plane
and a separated, extended, high-polarization region located along the
accretion disk axis, i.e., in a polar region. The whole system is
observed at intermediate- to high-inclination (Ogle \cite{ogl97}; Lamy and
Hutsem\'ekers \cite{lam04}; Smith et al. \cite{smi04}). The model is
schematically illustrated in Fig.~\ref{fig:model}.

These results depend little on the exact values of $\mu$ (which may be
lower at the I band wavelength) and of $F_{e} / F_{c}$.  A significant
($\gtrsim$ 0.2) contribution from the non-microlensed continuum is
nevertheless required to keep the polarization $p_{e}$ at reasonable
levels and to avoid fine-tuning, in agreement with the results of
Paper~II. These values of $F_{e} / F_{c}$ are compatible with
the fraction of polar scattered flux computed from models (e.g.,
Brown and McLean \cite{bro77}; Goosmann and Gaskell \cite{goo07}).

Time variability can be due to changes of the polarization or to a
variation of the $F_{e} / F_{c}$ ratio.  Interestingly, with $p_{c}$,
$\theta_{c}$, $p_{e}$, $\theta_{e}$ fixed, adopting a higher $F_{e} /
F_{c}$ ratio in 2011 than at previous epochs, as suggested in
Paper~II, rotates all polarization angles toward $\theta_{e}$ and
decreases the polarization degree in image D, in (qualitative)
agreement with the observations (Table~\ref{tab:pola}).

\begin{table}[t]
\caption{Results from the polarization model applied to the (D,A) pair}
\label{tab:polamod1d}
\begin{tabular}{lccccc}\hline\hline \\[-0.10in]
Epoch  / Band & $F_{e}/F_{c}$ & $p_{c} \;(\%)$ & $\theta_{c} \;(\degr)$ & 
$p_{e} \;(\%)$ & $\theta_{e} \;(\degr)$ \\ 
\hline \\[-0.10in]
1999 / F555W &   0.2 &   5.6$\pm$2.2 &  111$\pm$11 &  27.3$\pm$15.  & 32$\pm$16 \\
         &   0.5 &   6.3$\pm$2.3 &  112$\pm$12 &  12.5$\pm$7.3  & 34$\pm$17 \\
         &   0.8 &   7.1$\pm$3.0 &  113$\pm$12 &   8.9$\pm$5.3  & 35$\pm$17 \\
2008 / V &   0.2 &   3.5$\pm$0.5 &  107$\pm$4  &  16.6$\pm$3.2  & 31$\pm$6  \\
         &   0.5 &   3.9$\pm$0.6 &  108$\pm$4  &   7.7$\pm$1.5  & 34$\pm$6  \\
         &   0.8 &   4.3$\pm$0.6 &  109$\pm$4  &   5.5$\pm$1.1  & 36$\pm$6  \\
2011 / V &   0.2 &   1.7$\pm$0.4 &  121$\pm$8  &  16.9$\pm$3.0  & 44$\pm$5  \\
         &   0.5 &   2.0$\pm$0.5 &  124$\pm$7  &   8.6$\pm$1.4  & 46$\pm$5  \\
         &   0.8 &   2.4$\pm$0.6 &  125$\pm$7  &   6.5$\pm$1.1  & 46$\pm$5  \\
2011 / I &   0.2 &   2.1$\pm$1.0 &  105$\pm$13 &  17.4$\pm$6.9  & 31$\pm$11  \\
         &   0.5 &   2.5$\pm$1.2 &  107$\pm$13 &   8.7$\pm$3.3  & 32$\pm$11  \\
         &   0.8 &   2.9$\pm$1.3 &  109$\pm$13 &   6.5$\pm$2.4  & 33$\pm$11  \\
\hline\\[-0.2cm]
\end{tabular}
\end{table}

\subsection{The polarization in the line profiles}

The increase of the polarization degree in the BAL troughs and the
presence of shallower BALs in the polarized flux
(Fig.~\ref{fig:spola}) suggest that the extended polarized continuum
is absorbed by the BAL flow but less than the compact continuum (e.g.,
Goodrich and Miller \cite{goo95}). In other words, when the light from
the compact continuum is blocked by the absorber, the polarization
properties of the extended continuum are revealed in the BAL troughs.

\begin{figure}[t]
\resizebox{\hsize}{!}{\includegraphics*{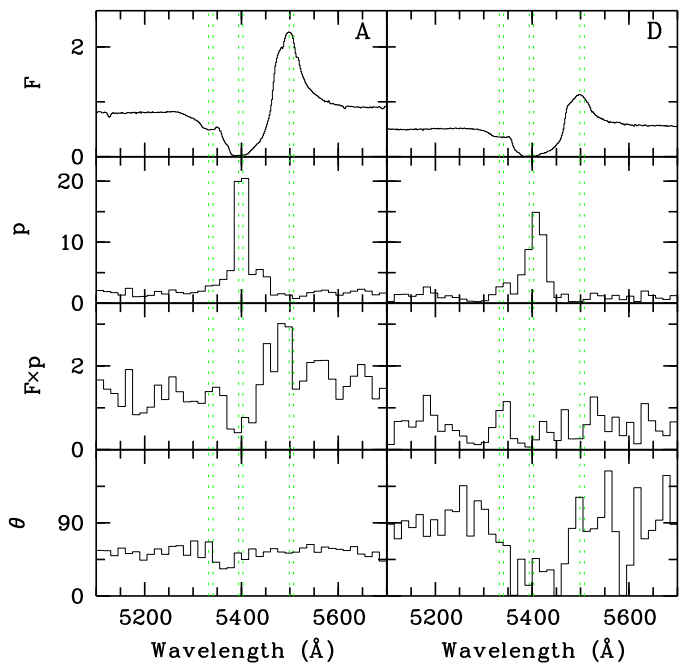}}
\caption{The polarization of the \ion{C}{iv} line.  From top to
bottom, the total flux $F(\lambda)$, (in arbitrary units), the linear
polarization degree $p(\lambda)$, (in \%), the polarized flux
$F(\lambda)\times p(\lambda)$, (in arbitrary units), and the
polarization position angle $\theta(\lambda)$ (in degree). The left
panel refers to image~A and the right panel to image~D. The position
of the broad emission line, the low-velocity BAL and the high-velocity
BAL are indicated.}
\label{fig:polac4}
\end{figure}

In the following we focus on the \ion{C}{iv} line, which shows the
clearest profiles (Fig.~\ref{fig:polac4}). In the BAL trough of image
A, the polarization angle rotates to $\sim$ 30 - 40\degr , which is
precisely the polarization angle derived for the polar-scattered
continuum. Moreover, the polarization in the troughs reaches high
degrees in agreement with the high values of $p_{e}$ derived from the
microlensing analysis (Table~\ref{tab:polamod1d}).  The rotation of
the polarization angle essentially occurs on the blue side of the
polarization maximum, a behavior also observed in images B and C.

In image D, microlensing magnifies the contribution of the compact
polarized continuum by a factor $\sim$ 2 with respect to image A,
which results in a rotation of the continuum polarization angle from
$\sim$ 55\degr\ to $\sim$ 85\degr. In the BAL trough, the light from
the compact source is blocked and the polarization angle jumps to the
polarization angle of the polar-scattered continuum, i.e.,
$\sim$~35\degr .  The polarization angle rotates over the 5310-5470
\AA\ wavelength range, which corresponds to the full intrinsic
absorption line extracted in Paper~I.

The emission lines are not significantly affected by microlensing
(Papers I and II).  They should then arise from a region larger than
the compact continuum (larger than 20 light days according to
O'Dowd et al. \cite{odo15}).  Since they appear in the polarized flux
and show a polarization degree lower than the continuum, they should
come from a region roughly cospatial with the extended scattering
region, as their lower polarization is due to geometric dilution
(Goodrich and Miller \cite{goo95}). Weakly polarized emission lines
dilute the continuum polarization and only slightly affect its
polarization angle.  In image D, the compact continuum is magnified
and the relative contribution of the emission lines is smaller.

\section{Discussion}

Our observations suggest the existence of two scattering
regions producing roughly perpendicularly polarized continua.  The
microlensed polarized continuum comes from a region close to the
accretion disk in the equatorial plane producing polarization
predominantly parallel to the accretion disk axis, while the
non-microlensed one comes from an extended polar region producing
perpendicular polarization.  In this scenario, we would expect the
accretion disk axis and thus the radio jet to have position angles
close to $\sim$ 115\degr.  Unfortunately, no clear information on the
object morphology is available\footnote{Kayser et al. (\cite{kay90})
have resolved the radio structure of \obj . However deformation by the
lens precluded the determination of the position angle of the
jet. Venturini and Solomon (\cite{ven03}) resolved the CO emission
from the source and measured its position angle, 55\degr , after
correcting for lens distortion. They refer to this structure as ``an
extended rotating molecular starburst disk''.  The position angle
inferred for the accretion disk axis ($\sim$115\degr ) is comparable
within the uncertainties to the position angle of this CO disk axis
($\sim$145\degr ).  However, the alignment of the supermassive
black hole, accretion disk, and host galaxy axes is still an open
question (e.g., Lagos et al. \cite{lag11}; Hopkins et
al. \cite{hop12}; Dubois et al. \cite{dub14}).}

Since an equatorial BAL wind can originate from the accretion
disk (Murray et al. \cite{mur95}), the polarization of the compact
continuum could be due to electron scattering at the base of
the wind, as proposed by Wang et al.~(\cite{wan05,wan07}). Observed at
high inclination, this ring-like region predominantly produces
parallel polarization. Both the central unpolarized source and this
compact scattering region are microlensed in image D.  On the other
hand, the extended polar scattering region could be related to polar
BAL and/or BEL regions (e.g., Ogle~\cite{ogl97}; Korista and Ferland
\cite{kor98}; Ghosh and Punsly \cite{gho07}), in agreement with the
two-component wind structure proposed in Paper~I (see also Borguet and
Hutsem\'ekers \cite{bor10}, and Borguet \cite{bor09}).  Finally,
resonance scattering in the equatorial and polar BAL flows could fill
in the red part of the absorption troughs with light polarized at the
same position angle as the continuum, so that a significant rotation
of the polarization angle is only observed on the blue side of the
main absorption trough. Detailed modeling would be needed to test
these hypotheses.

In Paper~II we found that, in 2011, the highest velocity BAL trough
absorbs the extended continuum more strongly than the compact
continuum. In that case, we would expect the polarization angle to
jump toward the value of the compact continuum, i.e., $\sim$ 115\degr
, in that trough. This is not observed (Fig.~\ref{fig:polac4}).  This
may indicate that the part of the extended scattering region that
contributes the most to the polarization differs from the part of the
extended region that suffers strong absorption in the highest velocity
BAL trough.  Alternatively, the high-velocity part of the wind can be
more clumpy than the low-velocity part.  The secular decrease of the
high-velocity absorption trough (Paper I and II) could be due to
clouds moving out of the line of sight (Capellupo et
al. \cite{cap14}), thus affecting the light from the compact region
more strongly than the light from the extended region.  In that case,
absorbing clouds could preferentially uncover the part of the
ring-like compact polarization source that produces polarization
perpendicular to the system axis, so that the polarization angle in
the high-velocity trough remains roughly unchanged.

\section{Conclusions}

We have obtained spectropolarimetric observations of the four images
of the gravitationally lensed BAL quasar H1413+117.  We found that the
polarization of image D, which is microlensed, is significantly
different from the polarization of image A, which is essentially
unaffected by microlensing.

The observations can be consistently interpreted by assuming that the
polarized continuum comes from two regions, producing roughly
perpendicular polarizations: a compact one located in the equatorial
plane close to the accretion disk and an extended one located along
the polar axis.  Our results provide further evidence for the
existence of a region in \obj\ that is more extended than the
accretion disk and that significantly contributes to the observed
continuum. Our findings allow us to identify this as a
polar-scattering region.

\begin{acknowledgements}
We thank M. Kishimoto for useful discussions and the referee for
constructive comments.  Support for T. Anguita is provided by the
Ministry of Economy, Development, and Tourism's Millennium Science
Initiative through grant IC120009, awarded to The Millennium Institute
of Astrophysics, MAS and proyecto FONDECYT 11130630. D. Sluse
acknowledges support from a {\it {Back to Belgium}} grant from the
Belgian Federal Science Policy (BELSPO), and partial funding from the
Deutsche Forschungsgemeinschaft, reference SL172/1-1.
\end{acknowledgements}

\begin{appendix}

\section{A simple model for the polarization of the continuum}

We consider two sources of continuum polarization. The compact source ($c$) is
microlensed by a factor $\mu$ and the extended source ($e$) is not
microlensed. Both are macrolensed by a factor $M$. We write for images 1
and~2
\begin{eqnarray} 
F_{1} & = & M \, F_{e} + M \mu \, F_{c} \; ,\\
S_{1} & = & M \, S_{e} + M \mu \, S_{c}\; ,\\
F_{2} & = & F_{e} + F_{c} \; ,\\
S_{2} & = & S_{e} + S_{c} \; ,
\end{eqnarray} 
where $S$ represents the Stokes fluxes $Q$ or $U$.  We assumed that
the compact source is magnified by the microlensing caustic network
without spatial distortions modifying its polarization.
Denoting $\beta$ = $F_{e} / F_{c}$, the flux
ratio of the extended and the compact continua, we have
\begin{eqnarray} 
s_{1} & = & \frac{\mu \, s_{c} + \beta \, s_{e}}{\mu+\beta} \; ,\\
s_{2} & = & \frac{s_{c} + \beta \, s_{e}}{1+\beta} \; ,
\end{eqnarray} 
where $s$ denotes the normalized Stokes parameters $q$ or $u$. These
equations can be solved to express the polarization of the compact
and extended continua as a function of the polarization
measured in images 1 and 2, i.e.,
\begin{eqnarray}
s_{e} & = & \frac{\mu \, (1+\beta) \, s_{2} - (\mu+\beta) \, 
                                            s_{1}} {\beta\,(\mu-1)} \; , \\
s_{c} & = &  \frac{(\mu+\beta) \, s_{1} - (1+\beta) \, s_{2}}{\mu-1} \; .  
\end{eqnarray} 
The polarization degrees and polarization angles can finally be
computed using $p_{x} = \sqrt{q_{x}^2+u_{x}^2}$ and $\theta_{x} = 1/2 \,
\arctan \,(u_{x}/q_{x})$, where $x$ = $c$ or $e$ for either the compact or
the extended continuum.
\end{appendix}
\end{document}